\documentstyle[amstex,preprint,aps]{revtex}

\begin{document}
\title{Two Qubit Quantum Computing in a Projected Subspace}
\author{Bi Qiao$^{abc}$, H. E. Ruda$^{a}$ and M. S. Zhan$^{b}$}
\address{$^{a}$Energenius Centre for Advanced Nanotechnology, University of\\
Toronto, Canada M5S 3E4.\\
$^{b}$Wuhan Institute of Physics and Mathmatics, Chinese Academy of Science,%
\\
Wuhan 430071, China.\\
$^{c}$Complexity Science Center, Yangzhou University, Yangzhou 225002, China.%
}
\maketitle

\begin{abstract}
A formulation for performing quantum computing in a projected subspace is
presented, based on the subdynamical kinetic equation (SKE) for an open
quantum system. The eigenvectors of the kinetic equation are shown to remain
invariant before and after interaction with the environment. However, the
eigenvalues in the projected subspace exhibit a type of phase shift to the
evolutionary states. This phase shift does not destroy the decoherence-free
(DF) property of the subspace because the associated fidelity is $1$. This
permits a universal formalism to be presented - the eigenprojectors of the
free part of the Hamiltonian for the system and bath may be used to
construct a DF projected subspace based on the SKE. To eliminate possible
phase or unitary errors induced by the change in the eigenvalues, a
cancellation technique is proposed, using the adjustment of the coupling
time, and applied to a two qubit computing system. A general criteria for
constructing a DF projected subspace from the SKE is discussed. Finally, a
proposal for using triangulation to realize a decoherence-free subsystem
based on SKE is presented. The concrete formulation for a two-qubit\ model
is given exactly. Our approach is novel and general, and appears applicable
to any type of decoherence.{\LARGE \ }

Key Words: Quantum Computing, Decoherence, Subspace, Open System

PACS number: $03.67.Lx,33.25.+k,.76.60.-k$

{\LARGE \ }

{\Large \ }

{\LARGE \ }
\end{abstract}

\section{Introduction}

Solid-state proposals for quantum computers have progressed markedly
recently: these include superconducting junctions, quantum dots, electron
spin resonance, nuclear spins of impurity atoms, and nuclear spins in a
crystal lattice$^{\left[ 1-5\right] }$. The working conditions for most
proposals require low operating temperatures (typically a few $K$). Ideally,
if such system can be created as isolated from the environment, system
evolution for a given quantum computing process, may be described by a
unitary operator with time reversal symmetry. Often this is not the case for
real quantum computing systems because the interaction between such systems
and their environment introduce decoherence, which destroys the
superpositions of qubits that enable the quantum logical operations to be
validated$^{\left[ 6,7\right] }$. Such decoherence is a major obstacle in
developing a practical quantum computer. Recent publications have formulated
a theory for a decoherence-free (DF) subspace in which quantum computing is
performed. The first formulation of DF conditions was performed by Paolo
Zanardi and Mario Rasetti$^{\left[ 8,9\right] }$. This original formulation
is quite general, which is not in the form of a master equation and does not
invoke the Born-Markov approximation. Further work by Viola and Zanardi deal
with DF subspaces by quantum control theory in a non-Markovian setting$^{%
\left[ 10,11,12\right] }$. Then an important direction of the DF subspace
theory was developed from a Master equation (such as the Lindblad equation)
for the open system within the powerful semigroup approach$^{\left[ 13-17%
\right] }$. A quantum jump description for the state of the atoms has also
derived a consistent result with Master equation for dipole interaction
atoms, but may give more insight into the time evolution of a single system$%
^{\left[ 18\right] }$. Experimentally, DF subspace have recently been
observed, which shows that such DF subspaces do indeed exist, allowing
logical qubits to be encoded without decoherence$^{\left[ 19,20\right] }$.
These provide the motivation for building appropriate open quantum computing
systems suitable for $\left( a\right) $ canceling the effects of intrinsic
decoherence, and thus behaving as ideal quantum computing systems in the
subspaces, or $\left( b\right) $ construct DF subspace without invoking any
approximations (such as the Born-Markov approximation) or restrictions on
the type of decoherence (e.g., symmetric and collective decoherence). In
this open system, self-adjoint operators and unitary evolution groups are
not intrinsically required for quantum computation. Quantum computation can
then be performed in a more general functional space including triplet
structures: Rigged Hilbert Space (RHS), $\Phi \subset {\cal H}\subset \Phi
^{\times }$, or Rigged Liouville Space (RLS), $\Phi \otimes \Phi \subset 
{\cal H}\otimes {\cal H}\subset $ $^{\times }\Phi \otimes \Phi ^{\times }$,
where ${\cal H}$ is Hilbert space, $\Phi $ is dense subspace of ${\cal H}$
and $\Phi ^{\times }$ is dual space of $\Phi $, rather than just Hilbert or
Liouville Space$^{\left[ 21-23\right] }$.

In this work, we present a formulation for performing quantum computation in
a DF projected subspace based on the subdynamical equation (SKE). In section
II, we briefly introduce a subdynamical formulation in the Hamiltonian
representation. In section III, we propose a type of DF projected subspace.
In sections IV and V, these concepts are applied to the example of a two
qubit spin computing system plus Bosonic bath; a procedure to cancel the
phase error induced by the change of eigenvalues is discussed. In section
VI, a general approach for obtaining the necessary and sufficient condition
for DF behavior is discussed. In section VII, a triangulating method is
introduced for determining the DF projected subspace exactly.

\section{Subdynamics Formulation}

Consider a quantum system which is composed of $N$ quantum registers and
interacts with a large thermal reservoir. We denote $H_{S}\left( t\right) $, 
$H_{B}$ and $H_{int}$ as the Hamiltonian of the system $S$, the Hamiltonian
of the the thermal reservoir $B$, and the interaction between $S$ and $B$,
respectively. Then the total Hamiltonian of the system plus the reservoir
can be expressed as $H_{S}\left( t\right) \otimes I_{B}+I_{S}\otimes
H_{B}+\lambda H_{int}$, with the corresponding Schr\"{o}dinger equation\ as $%
i\frac{\partial }{\partial t}\psi \left( t\right) \equiv H\left( t\right)
\psi \left( t\right) $. If one chooses the time-independent eigenprojectors
of $H_{S}\left( t\right) \otimes I_{B}+I_{S}\otimes H_{B}$ as $P_{\nu
}\equiv \left| \varphi _{\nu }\right\rangle \left\langle \varphi _{\nu
}\right| $ and $Q_{\nu }$ with $P_{\nu }$ + $Q_{\nu }$ = $1$ and the
eigenprojectors of the total Hamiltonian as $\Pi _{\nu }\left( t\right) $ to
satisfy the Heisenberg equation, then, from the definition of $\Pi _{\nu
}\left( t\right) $ one can induce the definition of the creation correlation
operator $C_{\nu }\left( t\right) $ ($D_{\nu }\left( t\right) $) as $Q_{\nu
}\Pi _{\nu }\left( t\right) =C_{\nu }\left( t\right) \Pi _{\nu }\left(
t\right) $ ($\Pi _{\nu }\left( t\right) Q_{\nu }=\Pi _{\nu }\left( t\right)
D_{\nu }\left( t\right) $), respectively. Note that the operator $C_{\nu
}\left( t\right) $ creates the $Q_{\nu }$-part of $\Pi _{\nu }\left(
t\right) $ from $P_{\nu }$ and the operator $D_{\nu }\left( t\right) $
destroys the $Q_{\nu }$-part of $\Pi _{\nu }\left( t\right) $ from $P_{\nu }$%
, since $C_{\nu }\left( t\right) =Q_{\nu }C_{\nu }\left( t\right) P_{\nu }$
and $D_{\nu }\left( t\right) =P_{\nu }D_{\nu }\left( t\right) Q_{\nu }$.
This enables a projected kinetic equation to be constructed in the projected
subspace by 
\begin{equation}
i\frac{\partial }{\partial t}\sum_{\nu }\left( P_{\nu }\Pi _{\nu }\left(
t\right) \psi \left( t\right) \right) =\Theta \left( t\right) \sum_{\nu
}\left( P_{\nu }\Pi _{\nu }\left( t\right) \psi \left( t\right) \right) ,
\label{eqn5}
\end{equation}%
with an intermediate operator defined as 
\begin{eqnarray}
\Theta \left( t\right) &\equiv &\left( H_{0}^{S}\left( t\right)
+H_{0}^{B}+H_{0}^{int}\right) +\left( H_{1}^{S}\left( t\right)
+H_{1}^{B}+H_{1}^{int}\right) C\left( t\right)  \label{eqn6} \\
&\equiv &\Theta _{0}\left( t\right) +H_{0}^{int}+\left( H_{1}^{S}\left(
t\right) +H_{1}^{B}+H_{1}^{int}\right) C\left( t\right) ,  \nonumber
\end{eqnarray}%
here $H_{0}^{S,B,int}\left( t\right) $ and $H_{1}^{S,B,int}\left( t\right) $
are the diagonal and off-diagonal parts of the corresponding total
Hamiltonian, respectively. $\Pi _{\nu }\left( t\right) $ can be found from
the subdynamics formulation as $\left( P_{\nu }+C_{\nu }\left( t\right)
\right) \left( P_{\nu }+D_{\nu }\left( t\right) C_{\nu }\left( t\right)
\right) ^{-1}\left( P_{\nu }+D_{\nu }\left( t\right) \right) ^{\left[ 21,22%
\right] }$. The creation (destruction) operator can be obtained from the
basic operator equations in the subdynamics formulation. Indeed, from the
definition of the eigenprojectors $\Pi _{\nu }\left( t\right) $ we have 
\begin{equation}
\left( E_{\nu }\left( t\right) -Q_{\nu }H\left( t\right) Q_{\nu }\right)
Q_{\nu }\Pi _{\nu }\left( t\right) =Q_{\nu }H_{int}P_{\nu }\Pi _{\nu }\left(
t\right) ,  \label{eqn7}
\end{equation}%
and hence yielding the crucial relationship 
\begin{equation}
Q_{\nu }\Pi _{\nu }\left( t\right) =Q_{\nu }C\left( t\right) P_{\nu }\Pi
_{\nu }\left( t\right) ,  \label{eqn8}
\end{equation}%
with definitions 
\begin{equation}
C\left( t\right) \equiv \sum_{\mu }\frac{1}{E_{\mu }\left( t\right) -Q_{\mu
}H\left( t\right) Q_{\mu }}Q_{\mu }H\left( t\right) P_{\mu },  \label{eqn9a}
\end{equation}%
and 
\begin{eqnarray}
C_{\nu }\left( t\right) &\equiv &Q_{\nu }C\left( t\right) P_{\nu }
\label{eqn9b} \\
&=&\frac{1}{E_{\nu }\left( t\right) -Q_{\nu }H\left( t\right) Q_{\nu }}%
Q_{\nu }H_{int}P_{\nu },  \nonumber
\end{eqnarray}%
where $E_{\nu }\left( t\right) $ is $\nu $th eigenvalue of the total
Hamiltonian $H\left( t\right) $. In the same way, 
\begin{equation}
D_{\nu }\left( t\right) =P_{\nu }H_{int}Q_{\nu }\frac{1}{E_{\nu }\left(
t\right) -Q_{\nu }H\left( t\right) Q_{\nu }}.  \label{eqn8b}
\end{equation}%
If one defines a projected wavefunction, $\psi ^{proj}\left( t\right) $, by $%
\psi \left( t\right) \equiv $ $%
\mathrel{\mathop{\sum }\limits_{\nu }}%
\left( P_{\nu }+C_{\nu }\left( t\right) \right) \psi _{\nu }^{proj}\left(
t\right) $, then $\sum_{\nu }P_{\nu }\Pi _{\nu }\left( t\right) \psi \left(
t\right) \equiv \sum_{\nu }\psi _{\nu }^{proj}\left( t\right) \equiv \psi
^{proj}\left( t\right) $, giving the formal evolution equation for a
projected wavefunction $\psi ^{proj}\left( t\right) $ as

\begin{eqnarray}
\psi ^{proj}\left( t\right) &=&\widehat{T}e^{-i\int_{t_{0}}^{t}dt^{\prime
}\Theta \left( t^{\prime }\right) }\psi ^{proj}\left( t\right) \left(
t_{0}\right)  \label{eqn9} \\
&=&\widehat{T}e^{-i\int_{t_{0}}^{t}dt^{\prime }\Theta \left( t^{\prime
}\right) }\sum_{\nu }\left( P_{\nu }+D_{\nu }\left( t_{0}\right) C_{\nu
}\left( t_{0}\right) \right) ^{-1}\left( P_{\nu }+D_{\nu }\left(
t_{0}\right) \right) \psi \left( t_{0}\right) .  \nonumber
\end{eqnarray}

\section{DF Project Subspace}

Using the above subdynamics formulation, it is apparent that before the
interaction occurs, the spectral decomposition for $\left( H_{S}\left(
t\right) +H_{B}\right) $ is the same as that for the total intermediate
operator $\Theta \left( t\right) $, i.e.,%
\begin{equation}
H_{S}\left( t\right) +H_{B}=\Theta \left( t\right) =\sum_{\nu }E_{\nu
}^{0}\left( t\right) P_{\nu },  \label{qn10}
\end{equation}%
where $E_{\nu }^{0}$ $\left( t\right) $ is the $\nu $th eigenvalue of $%
\left( H_{S}\left( t\right) +H_{B}\right) $. But once the interaction with
the environment has occurred, the eigenvalues change from $E_{\nu
}^{0}\left( t\right) $ to $E_{\nu }^{0}\left( t\right) +\triangle E_{\nu
}\left( t\right) $, where the definition%
\begin{equation}
\triangle E_{\nu }\left( t\right) =\left\langle \phi _{\nu }\right|
H_{1}^{int}C\left( t\right) \left| \phi _{\nu }\right\rangle ,  \label{eqn11}
\end{equation}%
is made. The corresponding eigenprojectors $P_{\nu }$ of $\Theta \left(
t\right) $ still remain invariant. The spectral decomposition of the
intermediate operator $\Theta \left( t\right) $ thus is different from $%
\left( H_{S}\left( t\right) +H_{B}\right) $, 
\begin{equation}
\Theta \left( t\right) =\sum_{\nu }\left( E_{\nu }^{0}\left( t\right)
+\triangle E_{\nu }\left( t\right) \right) P_{\nu }.  \label{eqn12a}
\end{equation}%
The corresponding mixed-state fidelity in the projected subspace can be
calculated from%
\begin{eqnarray}
F\left( t\right) &=&Tr\sqrt{\rho ^{proj}\left( t_{0}\right) \rho
^{proj}\left( t\right) }  \label{eqn13a} \\
&=&Tr\sqrt{\rho ^{proj}\left( t_{0}\right) \widehat{T}e^{-i\int \Theta
\left( t^{\prime }\right) dt^{\prime }}\rho ^{proj}\left( t_{0}\right) 
\widehat{T}e^{i\int \Theta \left( t^{\prime }\right) dt^{\prime }}} 
\nonumber \\
&=&Tr\sqrt{\left( \sum_{\nu ,\mu }\left( \rho ^{proj}\left( t_{0}\right)
\right) _{\nu \mu }P_{\nu \mu }\right) \left( \widehat{T}e^{-i\int \Theta
\left( t^{\prime }\right) dt^{\prime }}\sum_{\nu ,\mu }\left( \rho
^{proj}\left( t_{0}\right) \right) _{\nu \mu }P_{\nu \mu }\widehat{T}%
e^{i\int \Theta \left( t^{\prime }\right) dt^{\prime }}\right) }  \nonumber
\\
&=&Tr\sqrt{\sum_{\nu ,\mu }\left[ \left( \rho ^{proj}\left( t_{0}\right)
\right) _{\nu \mu }\right] ^{2}e^{-i\int_{t_{0}}^{t}\left[ E_{\nu
}^{0}\left( t^{\prime }\right) +\triangle E_{\nu }\left( t^{\prime }\right) %
\right] dt^{\prime }}P_{\nu \mu }e^{i\int_{t_{0}}^{t}\left[ E_{\mu
}^{0}\left( t^{\prime }\right) +\triangle E_{\mu }\left( t^{\prime }\right) %
\right] dt^{\prime }}}  \nonumber \\
&=&\sum_{\nu }\sqrt{\left\langle \phi _{\nu }\right| \sum_{\nu ,\mu }\left[
\left( \rho ^{proj}\left( t_{0}\right) \right) _{\nu \mu }\right]
^{2}e^{-i\int_{t_{0}}^{t}\left[ E_{\nu }^{0}\left( t^{\prime }\right)
+\triangle E_{\nu }\left( t^{\prime }\right) \right] dt^{\prime }}P_{\nu \mu
}e^{i\int_{t_{0}}^{t}\left[ E_{\mu }^{0}\left( t^{\prime }\right) +\triangle
E_{\mu }\left( t^{\prime }\right) \right] dt^{\prime }}\left| \phi _{\nu
}\right\rangle }  \nonumber \\
&=&\sum_{\nu }\sqrt{\left[ \left( \rho ^{proj}\left( t_{0}\right) \right)
_{\nu \nu }\right] ^{2}\ }=Tr\left( \sqrt{\rho ^{proj}\left( t_{0}\right)
\rho ^{proj}\left( t_{0}\right) }\right) =1.  \nonumber
\end{eqnarray}%
where $\left( \rho ^{proj}\left( t_{0}\right) \right) _{\nu \mu }$ are the
matrix elements of the density operator $\rho ^{proj}\left( t_{0}\right) $.
This exposes an exciting result: there is no decoherence in the projected
subspace for states of the system since the eigenvectors remain invariant.
However, a change in the eigenvalues introduces a phase shift in the
evolution of the states, 
\begin{equation}
e^{-i\int E_{\nu }^{0}\left( \tau \right) d\tau }\left| \psi _{\nu
}^{proj}\left( t_{0}\right) \right\rangle \text{ \ }\underrightarrow{\text{%
After the interaction}}\text{ \ }e^{-i\int \left( E_{\nu }^{0}\left( \tau
\right) +\triangle E_{\nu }\left( \tau \right) \right) d\tau }\left| \psi
_{\nu }^{proj}\left( t_{0}\right) \right\rangle .  \nonumber
\end{equation}%
From these findings there are two conclusions:

(1) In general, for any system $S+B$, we can use the eigenprojectors of $%
H^{S}$ to construct a DF projected subspace in which the eigenprojectors
remain invariant before and after the interaction between $S$ and $B$, while
the eigenvalues induce a phase shift in the eigenstates. The encoded states
in the projected subspace are the projected states which are related to the
original states by $P_{\nu }\Pi _{\nu }\psi _{\nu }$. In particular, in this
projected subsystem the states useful for performing quantum computing are
the reduced projected states $\rho _{S}^{proj}\left( t\right) $. These can
be obtained by using a projection operator $Tr_{R}P_{\nu }\Pi _{\nu }$ in
Liouville space to act on the density operator $\rho $ of the total system.
Although $\rho _{S}^{proj}\left( t\right) $ is not a reduced density
operator $\rho _{S}\left( t\right) $, one can consider $\rho
_{S}^{proj}\left( t\right) $ to be a generalization of $\rho _{S}\left(
t\right) $ (i.e., when $\Pi _{\nu }\longrightarrow P_{\nu }$, $\sum_{\nu
}P_{\nu }\Pi _{\nu }\rightarrow 1$, $\rho _{S}^{proj}\left( t\right) $ = $%
Tr_{B}\sum_{\nu }P_{\nu }\Pi _{\nu }\rho \left( t\right) $ = $Tr_{B}\rho
\left( t\right) $ = $\rho _{S}\left( t\right) $). We argue that the
information encoded in this projected subspace can be measured because the
projected bases are orthogonal and distinguishable in Liouville space -
namely $\left( \rho _{\nu }^{proj,S}\right| \left. \rho _{\mu
}^{proj,S}\right) $ = $0$, for $\nu \neq \mu $, because of $P_{\mu }P_{\nu
}=0$ in Liouville space too.

(2) The phase shift induced by the eigenvalues may cost a unitary type error
for quantum computation. Although this sort of error may be eliminated by
developing standard quantum error correction schemes (such as multiqubit code%
$^{\left[ 24\right] }$), the error recovery is not easy since phase error
induced by the systematic phase shift may occur, at the same time, to many
different clusters (inducing a phase or bit-flip errors in the encoded
data). The increase in the number of phase errors for different qubits will
cause a fast increase in the number of clusters, which likely leads to
impractical implementations. On the other hand, the property of invariant
eigenvectors in subspace provides the possibility of eliminating the unitary
error by adjusting the appropriate time scale for the evolution operator to
remain invariant under certain conditions; e.g., by choosing the time delay $%
\triangle t$ to allow $\sum_{\nu }e^{-i\int_{0}^{t}E_{\nu }^{0}\left( \tau
\right) d\tau }P_{\nu }=\sum_{\nu }e^{-i\int_{0}^{t+\triangle t}\left(
E_{\nu }^{0}\left( \tau \right) +\triangle E_{\nu }\left( \tau \right)
\right) d\tau }P_{\nu }$. Below, a concrete application of these concepts is
presented.

\section{Computing System}

We consider a two qubit quantum computing system $S$, consisting of spins $%
{\bf S}_{1}$ and ${\bf S}_{2}$, such as those corresponding to two electrons
around two $^{31}P$ confined in a germanium/silicon heterostructures in an
electron spin-resonance transistor$^{\left[ 4\right] }$, or for two
electrons confined in two quantum dots$^{\left[ 2,3\right] }$. Ignoring the
influence of the environment, the Hamiltonian may be written using the
Heisenberg model as 
\begin{equation}
H_{S}\left( t\right) =J\left( t\right) {\bf S}_{1}\cdot {\bf S}_{2},
\label{eqn14}
\end{equation}%
where $J\left( t\right) $ is the time-dependent exchange coupling parameter
determined by the specific model considerations. In the case of spins of the
two electrons (e.g., confined in two vertically (laterally) coupled quantum
dots), $J$ is the difference in the energies of two-electrons ground state,
a spin singlet at zero magnetic field, and the lowest spin-triplet state; $J$
is also a function of the electric and magnetic field and the interdot
distance$^{\left[ 25,26\right] }$. Using\ the relationship between ${\bf S}%
_{1}\cdot {\bf S}_{2}$ and the square of the sum of ${\bf S}_{1}$ and ${\bf S%
}_{2}$, the eigenvalues and eigenvectors of ${\bf S}_{1}\cdot {\bf S}_{2}$
can be found from ${\bf S}_{1}\cdot {\bf S}_{2}=\frac{1}{2}\left( {\bf S}%
^{2}-\frac{3}{2}\right) $ by 
\begin{eqnarray}
E_{S}^{1} &=&\frac{1}{4}\text{ \ \ }\left\{ 
\begin{array}{c}
\left| \phi _{1}\right\rangle =\left| 11\right\rangle , \\ 
\left| \phi _{2}\right\rangle =\left| 00\right\rangle , \\ 
\left| \phi _{3}\right\rangle =\frac{1}{\sqrt{2}}\left( \left|
01\right\rangle +\left| 10\right\rangle \right) ,%
\end{array}%
\right.  \label{eqn15} \\
E_{S}^{2} &=&-\frac{3}{4}\text{\ \ \ \ \ \ \ }\left| \phi _{4}\right\rangle =%
\frac{1}{\sqrt{2}}\left( \left| 01\right\rangle -\left| 10\right\rangle
\right) .  \nonumber
\end{eqnarray}%
A quantum XOR gate is given by the sequence of operations$^{\left[ 2\right]
} $, $U_{XOR}=e^{i\left( \pi /2\right) S_{1}^{z}}e^{-i\left( \pi /2\right)
S_{2}^{z}}U_{sw}^{1/2}e^{i\pi S_{1}^{z}}U_{sw}^{1/2}$, where $U_{sw}$ is an
(ideal) swap operator and determined generally by an evolution operator $%
U_{s}\left( J\left( 0\right) \tau \right) \ $by adjusting the coupling time
between the two spins in the evolution of the system. For the particular
spin-spin coupling duration, $\tau _{s}$ where $\int_{0}^{\tau _{s}}J\left(
t\right) dt=\pi \left( \text{mod }2\pi \right) $, $U_{sw}=U_{s}\left( \pi
\right) $, the swap operator, is given by 
\begin{equation}
U_{sw}=e^{-i\int_{0}^{\tau _{s}}H_{S}\left( \tau \right) d\tau
}=\sum_{j=1}^{3}e^{-i\frac{\pi }{4}}\left| \phi _{j}\right\rangle
\left\langle \phi _{j}\right| +e^{i\frac{3\pi }{4}}\left| \phi
_{4}\right\rangle \left\langle \phi _{4}\right|  \label{eqn16}
\end{equation}%
and can exchange the quantum states of qubit 1 and 2.

If we consider the influence of the environment, the non-ideal action of the
swap operation must be considered because the influence of environment
introduces decoherence. To treat this decoherence, one needs to understand
the behavior of the evolution of the system. Here it is assumed that the
environment consists of a set of harmonic-oscillators whose Hamiltonian is
given by $H_{B}=\sum_{k}\omega _{k}b_{k}^{+}b_{k}$, and the Hamiltonian
coupling to the two qubit spin system is $H_{int}=\lambda \sum_{k}\left(
\sigma _{z}^{1}+\sigma _{z}^{2}\right) \left( g_{k}b_{k}^{+}+g_{k}^{\ast
}b_{k}\right) $, where $b_{k}^{+}$, $b_{k}$ are bosonic operators for the $k$%
th field mode, characterized by a generally complex coupling parameter $%
g_{k} $ which characterizes the case as being one of either independent or
collective decoherence$^{\left[ 27\right] }$. The Hamiltonian operator for
the total system is given by $H\left( t\right) =H_{S}\left( t\right)
+H_{B}+\lambda H_{int},$ and the corresponding Schr\"{o}dinger equation is $i%
\frac{\partial }{\partial t}\psi \left( t\right) =\left( H_{S}\left(
t\right) +H_{B}+H_{int}\right) \psi \left( t\right) $. Choosing the
time-independent eigenprojectors of $H_{S}\left( t\right) +H_{B}$ as $P_{\nu
}$ and $Q_{\nu }$ with $Q_{\nu }+P_{\nu }=1$, with 
\begin{equation}
P_{\nu }\equiv \left| \varphi _{\nu }\right\rangle \left\langle \varphi
_{\nu }\right| =\left| n_{1}\cdots n_{k}\cdots \right\rangle \left| \phi
_{j}\right\rangle \left\langle \phi _{j}\right| \left\langle \cdots
n_{k}\cdots n_{1}\right|  \label{eqn17}
\end{equation}%
for $\nu =\left( j,n_{1}\cdots n_{k}\cdots \right) $ and $j=1,\cdots ,4$; $%
n_{k}=1$,$\cdots $, respectively, then the eigenprojectors $\Pi _{\nu
}\left( t\right) $ for the total Hamiltonian $H\left( t\right) $ can be
written in terms of the Heisenberg equation and satisfy the usual properties
of projection operators.

\section{Code Correction in Subspace}

Using above subdynamics formulation, it is apparent that before the
interaction occurs, the spectral decomposition for $\left( H_{S}\left(
t\right) +H_{B}\right) $ is the same as that for the total intermediate
operator $\Theta \left( t\right) $, i.e.,%
\begin{eqnarray}
&&H_{S}\left( t\right) +H_{B}=\Theta \left( t\right)  \label{eqn18a} \\
&=&\sum_{j=1}^{4}\sum_{n_{1}..n_{k}\cdots }E_{j,n_{1}\cdots n_{k}\cdots
}^{0}\left( t\right) P_{j,n_{1}\cdots n_{k}\cdots },  \nonumber
\end{eqnarray}%
where $E_{j,n_{1}\cdots n_{k}\cdots }^{0}$ $\left( t\right) $ = $\left( 
\frac{1}{4}-\delta _{j4}\frac{3}{4}\right) J\left( t\right) +\sum_{k}\omega
_{k}n_{k}$. But once interaction with the environment occurs, the
eigenvalues change from $E_{j,n_{1}\cdots n_{k}\cdots }^{0}\left( t\right) $
to $E_{j,n_{1}\cdots n_{k}\cdots }^{0}\left( t\right) +\triangle
E_{j,n_{1}\cdots n_{k}\cdots }\left( t\right) $, where we have made the
following definitions:%
\begin{equation}
\triangle E_{j,n_{1}\cdots n_{k}\cdots }\left( t\right) =\left\langle
n_{1}\cdots n_{k}\cdots \right| \left\langle \phi _{j}\right|
H_{int}C_{j,n_{1}\cdots n_{k}\cdots }\left( t\right) \left| \phi
_{j}\right\rangle \left| n_{1}\cdots n_{k}\cdots \right\rangle ,
\label{eqn19a}
\end{equation}%
while the corresponding the eigenvectors of $\Theta \left( t\right) $, $%
\left| \phi _{j}\right\rangle \left| n_{1}\cdots n_{k}\cdots \right\rangle $
remain invariant. The spectral decomposition of the intermediate operator $%
\Theta \left( t\right) $ is thus different from $H_{0}\left( t\right) $, 
\begin{equation}
\Theta \left( t\right) =\sum_{j=1}^{4}\sum_{n_{1}\cdots n_{k}\cdots }\left(
E_{j,n_{1}\cdots n_{k}\cdots }^{0}\left( t\right) +\triangle
E_{j,n_{1}\cdots n_{k}\cdots }\left( t\right) \right) P_{j,n_{1}\cdots
n_{k}\cdots }.  \label{eqn20}
\end{equation}%
This shows that there is no decoherence in the projected subspace for the
stationary states of the system since the eigenvectors remain invariant, but
for the evolutionary states, the change of the eigenvalues can introduce a
type of unitary like error in the system evolution, 
\begin{eqnarray}
&&\left| \psi ^{proj}\left( t\right) \right\rangle  \label{eqn21} \\
&=&\sum_{j=1}^{4}\sum_{n_{1}\cdots n_{k}\cdots }e^{-i\int_{0}^{t}\left(
E_{j,n_{1}\cdots n_{k}\cdots }^{0}\left( \tau \right) +\triangle
E_{j,n_{1}\cdots n_{k}\cdots }\left( \tau \right) \right) d\tau }\left| \psi
_{j,n_{1}\cdots n_{k}\cdots }^{proj}\left( 0\right) \right\rangle . 
\nonumber
\end{eqnarray}

As mentioned previously, the quantum XOR operator can be constructed from a
sequence of operations related to the ideal swap operator, adjusted by
controlling the coupling time for the interaction between the two qubits
without any influence from the environment. But if the effects of the
environment are now included, the ideal swap operator changes to the
non-ideal swap operator, owing to the unitary error. To cancel this type of
decoherence, in terms of the subdynamics formulation, it is proposed that
one allow the quantum logical operators to work on the projected subspaces.
For example, if the quantum XOR operator previously introduced is
considered, the ideal swap operator should be adjusted by controlling the
coupling time between the two spins without considering interactions with
the environment; then the ideal swap operator is given in a projected
subspace as $U_{sw}\left( \tau _{s}\right) $ = $e^{-i\left( \pi {\bf S}%
_{1}\cdot {\bf S}_{2}\otimes I_{B}+I_{S}\otimes H_{B}\tau _{s}\right) }$,
where a specific coupling duration $\tau _{s}$ is given by $\int_{0}^{\tau
_{s}}J\left( \tau \right) d\tau =\pi $ $\left( 
\mathop{\rm mod}%
2\pi \right) $. The unitary error is related to the non-ideal action of the
swap operator which can be adjusted by the evolution operator\ in a
projected subspace as $U_{sw}^{\prime }\left( \tau _{s}+\triangle t\right) $
= $e^{-i\int_{0}^{\tau _{s}+\triangle t}d\tau \left[ J\left( \tau \right) 
{\bf S}_{1}\cdot {\bf S}_{2}\otimes I_{B}+I_{S}\otimes
H_{B}+H_{int}C_{j,n_{1}\cdots n_{k}\cdots }\left( \tau \right) \right] }$.
The spectral decomposition of $U_{sw}^{\prime }\left( \tau _{s}+\triangle
t\right) $ can be expressed by adjusting the interaction time $\tau _{s}$ to 
$\tau _{s}+\triangle t$, allowing the non-ideal swap operator to be equal to
the ideal swap operator, 
\begin{eqnarray}
&&U_{sw}^{\prime }\left( \tau _{s}+\triangle t\right) =U_{sw}  \nonumber \\
&=&\sum_{n_{1}\cdots n_{k}\cdots }\left\{ \sum_{j=1}^{3}e^{-i\int_{0}^{\tau
_{s}+\triangle t}d\tau \left[ \frac{1}{4}J\left( \tau \right)
+\sum_{k}\omega _{k}n_{k}+\triangle E_{j,n_{1}\cdots n_{k}\cdots }\left(
t\right) \right] }P_{j,n_{1}\cdots n_{k}\cdots }\right.  \label{eqn22} \\
&&\left. +e^{-i\int_{0}^{\tau _{s}+\triangle t}d\tau \left[ -\frac{3}{4}%
J\left( \tau \right) +\sum_{k}\omega _{k}n_{k}+\triangle E_{j,n_{1}\cdots
n_{k}\cdots }\left( t\right) \right] }P_{4,n_{1}\cdots n_{k}\cdots }\right\}
\nonumber \\
&=&\sum_{n_{1}\cdots n_{k}\cdots }\left( \sum_{j=1}^{3}e^{-i\frac{\pi }{4}%
-i\tau _{s}\sum_{k}\omega _{k}n_{k}}P_{j,n_{1}\cdots n_{k}\cdots }+e^{i\frac{%
3\pi }{4}-i\tau _{s}\sum_{k}\omega _{k}n_{k}}P_{4,n_{1}\cdots n_{k}\cdots
}\right) .  \nonumber
\end{eqnarray}%
This induces the integral equation for determining $\triangle t$ which
depends on $j,n_{1}\cdots n_{k}\cdots $, 
\begin{equation}
\int_{0}^{\tau _{s}+\triangle t}d\tau \left( \frac{J\left( \tau \right) }{4}%
-\delta _{j4}\frac{3J\left( \tau \right) }{4}+\triangle E_{j,n_{1}\cdots
n_{k}\cdots }\left( \tau \right) \right) =\left\{ 
\begin{array}{c}
\frac{\pi }{4}\text{, for }j=1,2,3, \\ 
-\frac{3\pi }{4}\text{, for }j=4.%
\end{array}%
\right.  \label{eqn23}
\end{equation}%
For instance, if the exchange interaction coupling $J\left( {\bf B,E,}%
d\right) $ is a time-independent function of the external magnetic field $%
{\bf B}$, electric field ${\bf E}$ and the interdot distance $d$ as in the
case of vertically or laterally tunnel-coupled quantum dots, then $\triangle
t$ can be solved from Eq.(\ref{eqn23}) as: 
\begin{equation}
\triangle t=-\frac{\tau _{s}}{\frac{J\left( {\bf B,E,}d\right) }{4\triangle
E_{j,n_{1}\cdots n_{k}\cdots }}+1}=-\frac{\tau _{s}}{\frac{-3J\left( {\bf %
B,E,}d\right) }{4\triangle E_{4,n_{1}\cdots n_{k}\cdots }}+1},  \label{eqn24}
\end{equation}%
where the concrete formula for $J\left( {\bf B,E,}d\right) $, in the case of
vertically or laterally tunnel-coupled quantum dots, can be found in refs.$%
\left[ 25,26\right] $. Assuming, in this case, that the energy of the total
system is uniformly distributed, i.e., 
\begin{equation}
\frac{J\left( {\bf B,E,}d\right) }{4\triangle E_{j,n_{1}\cdots n_{k}\cdots }}%
=\frac{-3J\left( {\bf B,E,}d\right) }{4\triangle E_{4,n_{1}\cdots
n_{k}\cdots }}=const\text{, for }j=1,2,3;n_{1},\cdots ,n_{k}=1,2,\cdots .
\label{eqn25a}
\end{equation}%
The shift energy can be obtained from Eq. (\ref{eqn19a}), 
\begin{equation}
\triangle E_{j,n_{1}\cdots n_{k}\cdots }=\sum_{j^{\prime
}=1}^{4}\sum_{n_{1}^{\prime }\cdots n_{k}^{\prime }\cdots }\left(
H_{int}\right) _{j,n_{1}\cdots n_{k}\cdots ;j^{\prime },n_{1}^{\prime
}\cdots n_{k}^{\prime }\cdots }C_{j^{\prime },n_{1}^{\prime }\cdots
n_{k}^{\prime }\cdots ;j,n_{1}\cdots n_{k}\cdots },  \label{eqn26a}
\end{equation}%
and the matrix of the creation operator is given by 
\begin{equation}
C_{j^{\prime },n_{1}^{\prime }\cdots n_{k}^{\prime }\cdots ;j,n_{1}\cdots
n_{k}\cdots }=\left\langle \varphi _{j^{\prime },n_{1}^{\prime }\cdots
n_{k}^{\prime }\cdots }\right| C_{j,n_{1}\cdots n_{k}\cdots }\left| \varphi
_{j,n_{1}\cdots n_{k}\cdots }\right\rangle ,  \label{eqn27}
\end{equation}%
which can be calculated from formula (\ref{eqn9b}); for example, the first
order $C_{j,n_{1}\cdots n_{k}\cdots }^{\left[ 1\right] }$ is divided by
degenerate part and non-degenerate part:%
\begin{eqnarray}
&&C_{j,n_{1}\cdots n_{k}\cdots }^{\left[ 1\right] }  \label{eqnww3} \\
&=&\sum_{n_{1}^{\prime }\cdots n_{k}^{\prime }\cdots }\left[ \sum_{j^{\prime
}\neq j}^{3}\frac{1}{\triangle E_{j,n_{1}\cdots n_{k}\cdots }^{\left[ 2%
\right] }}\left( H_{int}\right) _{j^{\prime },n_{1}^{\prime }\cdots
n_{k}^{\prime }\cdots ;j,n_{1}\cdots n_{k}\cdots }\right.  \nonumber \\
&&+\left. \delta _{j^{\prime }4}\frac{1}{E_{j,n_{1}\cdots n_{k}\cdots
}^{0}-E_{j^{\prime },n_{1}^{\prime }\cdots n_{k}^{\prime }\cdots }^{0}}%
\left( H_{int}\right) _{j^{\prime },n_{1}^{\prime }\cdots n_{k}^{\prime
}\cdots ;j,n_{1}\cdots n_{k}\cdots }\right] P_{j^{\prime },n_{1}^{\prime
}\cdots n_{k}^{\prime }\cdots ;j,n_{1}\cdots n_{k}\cdots }  \nonumber \\
&=&\left\{ 
\begin{array}{c}
\frac{1}{\triangle E_{j,n_{1}\cdots n_{k}\cdots }^{\left[ 2\right] }}%
\sum_{n_{1}^{\prime }\cdots n_{k}^{\prime }\cdots }\left( H_{int}\right)
_{j,n_{1}^{\prime }\cdots n_{k}^{\prime }\cdots ;j,n_{1}\cdots n_{k}\cdots
}P_{j,n_{1}^{\prime }\cdots n_{k}^{\prime }\cdots ;j,n_{1}\cdots n_{k}\cdots
}\text{, for }j=1,2, \\ 
0,\text{ for }j=3,4,%
\end{array}%
\right. ,  \nonumber
\end{eqnarray}%
which results in the second order $\triangle E_{j,n_{1}\cdots n_{k}\cdots }^{%
\left[ 2\right] }$ by Eq.(\ref{eqn26a}) 
\begin{equation}
\triangle E_{j,n_{1}\cdots n_{k}\cdots }^{\left[ 2\right] }=\left\{ 
\begin{array}{c}
\sqrt{\sum_{j=1}^{2}\sum_{n_{1}^{\prime }\cdots n_{k}^{\prime }\cdots
}\left( \left( H_{int}\right) _{j,n_{1}^{\prime }\cdots n_{k}^{\prime
}\cdots ;j,n_{1}\cdots n_{k}\cdots }\right) ^{2}}, \\ 
0,\text{ for }j=3,4,%
\end{array}%
\right.  \label{eqnww2}
\end{equation}%
with the matrix elements given by 
\begin{eqnarray}
\left( H_{int}\right) _{j,n_{1}\cdots n_{k}\cdots ;j^{\prime },n_{1}^{\prime
}\cdots n_{k}^{\prime }\cdots } &=&\ \left\langle n_{1}\cdots n_{k}\cdots
\right| \left\langle \phi _{i}\right| H_{int}\left| \phi _{j^{\prime
}}\right\rangle \left| n_{1}^{\prime }\cdots n_{k}^{^{\prime }}\cdots
\right\rangle  \label{qn17a} \\
&=&\left\{ 
\begin{array}{c}
\lambda \left( \delta _{1,j^{\prime }}-\delta _{2,j^{\prime }}\right) 
\mathrel{\mathop{\sum }\limits_{k}}%
\left( \delta _{n_{k},n_{k}^{\prime }+1}g_{k}\sqrt{n_{k}^{\prime }+1}+\delta
_{n_{k},n_{k}^{\prime }-1}g_{k}^{\ast }\sqrt{n_{k}^{\prime }}\right) , \\ 
0,\text{ for }j=3,4,%
\end{array}%
\right. .  \nonumber
\end{eqnarray}%
Eq.(\ref{eqn24}) shows that although the interaction introduces the sort of
phase shift in the swap operator, this sort of phase shift can be cancelled
by adjusting the coupling time between the two spins under the assumptions
of homogeneous distribution of energy (i.e., owing to the invariance of the
eigenvectors in the projected subspace). In the same way, the second-order
projected states in the projected subspace are given by 
\begin{eqnarray}
{} &&\psi _{j,n_{1}\cdots n_{k}\cdots }^{proj\left[ 2\right] }\left(
t\right) =\left( P_{j,n_{1}\cdots n_{k}\cdots }+D_{j,n_{1}\cdots n_{k}\cdots
}-D_{j,n_{1}\cdots n_{k}\cdots }C_{j,n_{1}\cdots n_{k}\cdots }\right) \left|
\psi \left( t\right) \right\rangle  \label{eqn18bb} \\
&=&\left\{ P_{j,n_{1}\cdots n_{k}\cdots }-\frac{\left( \delta _{1j}+\delta
_{2j}\right) }{\triangle E_{j,n_{1}\cdots n_{k}\cdots }^{\left[ 2\right] }}%
\sum_{n_{1}^{\prime }\cdots n_{k}^{\prime }\cdots }\left( H_{int}\right)
_{j,n_{1}\cdots n_{k}\cdots ;j,n_{1}^{\prime }\cdots n_{k}^{\prime }\cdots
}P_{j,n_{1}\cdots n_{k}\cdots ;j,n_{1}^{\prime }\cdots n_{k}^{\prime }\cdots
}\right.  \nonumber \\
&&\left. -\frac{\left( \delta _{1j}+\delta _{2j}\right) }{\left( \triangle
E_{j,n_{1}\cdots n_{k}\cdots }^{\left[ 2\right] }\right) ^{2}}%
\sum_{n_{1}^{\prime }\cdots n_{k}^{\prime }\cdots }\sum_{n_{1}^{\prime
\prime }\cdots n_{k}^{\prime \prime }\cdots }\left( \left( H_{int}\right)
_{j,n_{1}\cdots n_{k}\cdots ;j,n_{1}^{\prime }\cdots n_{k}^{\prime }\cdots
}\left( H_{int}\right) _{j,n_{1}^{\prime \prime }\cdots n_{k}^{\prime \prime
}\cdots ;j,n_{1}\cdots n_{k}\cdots }\right) P_{j,n_{1}\cdots n_{k}\cdots
}\right\} \left| \psi \left( t\right) \right\rangle ,  \nonumber
\end{eqnarray}%
where the first order destruction operator is%
\begin{eqnarray}
&&D_{j,n_{1}\cdots n_{k}\cdots }^{\left[ 1\right] }  \label{eqnbb56} \\
&=&\left\{ 
\begin{array}{c}
\sum_{n_{1}^{\prime }\cdots n_{k}^{\prime }\cdots }\frac{-1}{\triangle
E_{j,n_{1}\cdots n_{k}\cdots }^{\left[ 2\right] }}\left( H_{int}\right)
_{j,n_{1}\cdots n_{k}\cdots ;j,n_{1}^{\prime }\cdots n_{k}^{\prime }\cdots
}P_{j,n_{1}^{\prime }\cdots n_{k}^{\prime }\cdots ;j,n_{1}\cdots n_{k}\cdots
},\text{ for }j=1,2, \\ 
0,\text{ for }j=3,4,%
\end{array}%
\right. .  \nonumber
\end{eqnarray}%
By above cancelling procedure, the evolution formula for second-order
reduced projected density operator, before or after the interaction, remains
the same and is given by%
\begin{eqnarray}
&&\rho _{j,n_{1}\cdots n_{k}\cdots ;j^{\prime },n_{1}\cdots n_{k}\cdots
}^{proj\left[ 2\right] ,S}\left( t\right)  \label{eqnbcbc} \\
&=&Tr_{B}e^{-i\Theta _{j,n_{1}\cdots n_{k}\cdots }^{\left[ 2\right]
}t}p_{j,n_{1}\cdots n_{k}\cdots ;j^{\prime },n_{1}^{\prime }\cdots
n_{k}^{\prime }\cdots }\left| \varphi _{j,n_{1}\cdots n_{k}\cdots }^{proj%
\left[ 2\right] }\left( 0\right) \right\rangle \left\langle \varphi
_{j^{\prime },n_{1}^{\prime }\cdots n_{k}^{\prime }\cdots }^{proj\left[ 2%
\right] }\left( 0\right) \right| e^{i\Theta _{j^{\prime },n_{1}^{\prime
}\cdots n_{k}^{\prime }\cdots }^{\left[ 2\right] }t}  \nonumber \\
&=&e^{-i\left( E_{j}^{0}-E_{j^{\prime }}^{0}\right) t}\rho _{j,n_{1}\cdots
n_{k}\cdots ;j^{\prime },n_{1}\cdots n_{k}\cdots }^{proj,S}\left( 0\right) ,
\nonumber
\end{eqnarray}%
where $p_{j,n_{1}\cdots n_{k}\cdots ;j^{\prime },n_{1}^{\prime }\cdots
n_{k}^{\prime }\cdots }$ represent probability of existence for the state $%
\left| \varphi _{j,n_{1}\cdots n_{k}\cdots }^{proj\left[ 2\right] }\left(
0\right) \right\rangle \left\langle \varphi _{j^{\prime },n_{1}^{\prime
}\cdots n_{k}^{\prime }\cdots }^{proj\left[ 2\right] }\left( 0\right)
\right| $, noting that $\rho _{j,n_{1}\cdots n_{k}\cdots ;j^{\prime
},n_{1}\cdots n_{k}\cdots }^{proj,S}\left( 0\right) $ is given by initial
decoupling condition, $\rho _{j,n_{1}\cdots n_{k}\cdots ;j^{\prime
},n_{1}\cdots n_{k}\cdots }^{proj,S}\left( 0\right) =Tr_{B}\left(
P_{j,n_{1}\cdots n_{k}\cdots }\rho _{S}\left( 0\right) \otimes \rho
_{B}\left( 0\right) P_{j^{\prime },n_{1}^{\prime }\cdots n_{k}^{\prime
}\cdots }\right) $. The fidelity, which measures the decoherence between the
initial state $\psi ^{proj\left[ 2\right] }\left( t\right) $ and the
evolution reduced project density operator $\rho ^{proj\left[ 2\right]
,S}\left( t\right) $ in the subspaces, is given by%
\begin{eqnarray}
&&F\left( \left| \psi ^{proj\left[ 2\right] }\left( t\right) \right\rangle
,\rho ^{proj\left[ 2\right] ,S}\left( t\right) \right)  \label{eqn20e} \\
&=&\sqrt{\left\langle \psi ^{proj\left[ 2\right] }\left( t\right) \right|
\sum_{j,j^{\prime },n_{1}\cdots n_{k}\cdots }\rho _{j,n_{1}\cdots
n_{k}\cdots ;j^{\prime },n_{1}\cdots n_{k}\cdots }^{proj\left[ 2\right]
,S}\left( t\right) \left| \psi ^{proj\left[ 2\right] }\left( t\right)
\right\rangle }  \nonumber \\
&=&\sqrt{\left\langle \sum_{j,n_{1}\cdots n_{k}\cdots }\varphi
_{j,n_{1}\cdots n_{k}\cdots }^{proj\left[ 2\right] }\left( 0\right) \right|
\sum_{j,j^{\prime },n_{1}\cdots n_{k}\cdots }\rho _{j,n_{1}\cdots
n_{k}\cdots ;j^{\prime },n_{1}\cdots n_{k}\cdots }^{proj\left[ 2\right]
,S}\left( 0\right) \left| \sum_{j^{\prime },n_{1}\cdots n_{k}\cdots }\varphi
_{j^{\prime },n_{1}^{\prime }\cdots n_{k}^{\prime }\cdots }^{proj\left[ 2%
\right] }\left( 0\right) \right\rangle }  \nonumber \\
&=&\sqrt{\sum_{j,n_{1}\cdots n_{k}\cdots }p_{j,n_{1}\cdots n_{k}\cdots
}\sum_{j^{\prime },n_{1}\cdots n_{k}\cdots }p_{j^{\prime },n_{1}\cdots
n_{k}\cdots }}  \nonumber \\
&=&1,  \nonumber
\end{eqnarray}%
which shows that there is no decoherence after the cancellation process.

\section{The DF Projected Subspace}

A completely decoherence-free condition (i.e., including no phase shift) in
the projected subspace can also be determined, due to the definition of the
intermediate operator of Eq.(\ref{eqn6}). That is 
\begin{equation}
H_{int}C\left( t\right) =0.  \label{eqn18}
\end{equation}%
In this case, eigenvalues for the intermediate operators in the projected
subspaces, are the same as those for the total Hamiltonian in the total
space, while the eigenvectors are the same as those of the unperturbed
Hamiltonian. The projected subspaces spanned by the set of the eigenvectors
of the intermediate operator are invariant for the evolutionary and
stationary states of the system, and are closed for the intermediate
operators. This allows quantum computing with coherent states such that no
information leaks out from the projected subspaces. Hence one can use the
full power of DF subspaces to preserve coherence for full scale quantum
computing. The key question then is how to ensure that Eq. (\ref{eqn18})
holds. In the Born-Markov approximation ( e.g., Lindblad equation approach)
and the assumptions of symmetric or collective decoherence$^{\left[ 6,7,13,24%
\right] }$, if $P_{\mu \text{ }}$ are partial eigenprojectors of $H_{int}$
(i.e., $H_{int}P_{\mu \text{ }}=\varepsilon _{\alpha }P_{\mu \text{ }}$, $%
\forall $ degenerate $\alpha \in $ set of complex number $c$) then it means
having $P_{\nu }H_{int}P_{\mu }C_{\nu }\left( t\right) P_{\nu }=0$, for any $%
\nu $. But in the general case, without restrictions on the type of
decoherence (e.g., a non-Markovian process and non-symmetric and
non-collective decoherence) how can one find the conditions for constructing
the DF subspace? For generality, we propose a procedure to find the
condition for the DF subspace starting directly from Eq.(\ref{eqn18}) by
considering weak coupling between the system and the environment, without
making any of the above assumptions. In fact, from Eq.(\ref{eqn18}) we have:%
\begin{eqnarray}
&&P_{j,n_{1}\cdots n_{k}\cdots }H_{int}C_{j,n_{1}\cdots n_{k}\cdots }\left(
t\right)  \label{eqn19} \\
&=&\sum_{j^{\prime }=1}^{4}\sum_{n_{1}^{\prime }\cdots n_{k}^{\prime }\cdots
}\frac{1}{E_{j,n_{1}\cdots n_{k}\cdots }^{0}\left( t\right) -E_{j^{\prime
},n_{1}^{\prime }\cdots n_{k}^{\prime }\cdots }^{0}\left( t\right)
+\triangle E_{j,n_{1}\cdots n_{k}\cdots }\left( t\right) }P_{j,n_{1}\cdots
n_{k}\cdots }H_{int}P_{j^{\prime },n_{1}^{\prime }\cdots n_{k}^{\prime
}\cdots }H_{int}P_{j,n_{1}\cdots n_{k}\cdots }  \nonumber \\
&=&\frac{\lambda ^{2}}{E_{1,n_{1}\cdots n_{k}\cdots }^{0}\left( t\right)
-E_{1,n_{1}\cdots n_{k}^{\prime }\cdots }^{0}\left( t\right) +\triangle
E_{1,n_{1}\cdots n_{k}\cdots }\left( t\right) }\sum_{k}\left(
g_{k}g_{k+1}^{\ast }\left( n_{k}+1\right) +g_{k-1}g_{k}^{\ast }n_{k}\right)
\delta _{j1}P_{1,n_{1}\cdots n_{k}\cdots }  \nonumber \\
&&+\frac{\lambda ^{2}}{E_{2,n_{1}\cdots n_{k}\cdots }^{0}\left( t\right)
-E_{2,n_{1}\cdots n_{k}^{\prime }\cdots }^{0}\left( t\right) +\triangle
E_{2,n_{1}\cdots n_{k}\cdots }\left( t\right) }\sum_{k}\left(
g_{k}g_{k+1}^{\ast }\left( n_{k}+1\right) +g_{k-1}g_{k}^{\ast }n_{k}\right)
\delta _{2j}P_{2,n_{1}\cdots n_{k}\cdots }  \nonumber \\
&=&0.  \nonumber
\end{eqnarray}%
This gives a DF condition which is a restriction on the bath operator by 
\begin{equation}
\sum_{k}\left( g_{k}g_{k+1}^{\ast }\left( n_{k}+1\right) +g_{k-1}g_{k}^{\ast
}n_{k}\right) =0.  \label{eqn23a}
\end{equation}%
In one spectial case it is 
\begin{equation}
\frac{n_{k}+1}{n_{k}}=-\frac{g_{k-1}g_{k}^{\ast }}{g_{k}g_{k+1}^{\ast }}.
\label{eqnbv}
\end{equation}

Under the DF conditions, e.g. with Eqs.(\ref{eqn23a}) or (\ref{eqnbv})
holding, Eq.(\ref{eqn5}) reduces to%
\begin{equation}
i\frac{\partial \left| \varphi ^{proj}\left( t\right) \right\rangle }{%
\partial t}=H_{0}\left( t\right) \left| \varphi ^{proj}\left( t\right)
\right\rangle  \label{eqn25}
\end{equation}%
with evolution of the projected state described by%
\begin{eqnarray}
\left| \varphi ^{proj}\left( t\right) \right\rangle &=&\widehat{T}%
e^{-itH_{0}\left( t\right) }\left| \varphi ^{proj}\left( 0\right)
\right\rangle  \label{eqn26} \\
&=&\widehat{T}e^{-itH_{0}\left( t\right) }\sum_{\nu }\left( P_{\nu }+D_{\nu
}\left( 0\right) \right) ^{-1}\left( P_{\nu }+D_{\nu }\left( 0\right)
\right) \left| \varphi \left( 0\right) \right\rangle ,  \nonumber
\end{eqnarray}%
which enables the constructed projected subspaces to become a DF subspace
spanned by the set of $\left\{ P_{j,n_{1}\cdots n_{k}\cdots }\right\} $,
although the total space is subject to decoherence. The projected subspace
is closed with respect to the intermediate operator, $\Theta _{\nu }\left(
t\right) \Phi ^{proj}\subset \Phi ^{proj}$, for any projected state in the
projected subspace $\Phi ^{proj}$, $\left( P_{\nu }\Pi _{\nu }\left(
t\right) \psi \left( t\right) \right) $. Furthermore, if one generalizes the
initial test projector $P_{\mu \text{ }}$to be up (or down) triangular from
the partial eigenprojector of $H_{int}$, i.e. $P_{\nu }H_{int}P_{\mu \text{ }%
}\neq 0$ for $\nu <\mu $, otherwise it is zero, then Eq.(\ref{eqn6}) gives $%
C_{\nu }\left( t\right) P_{\nu }=0$ which enables Eq.(\ref{eqn18}) to
generally hold in the projected subspace without any approximation. In this
sense, if one can construct a triangular basis for $H_{int}$ to span a
projected subspace, then this space is DF although the total space is
decoherent.

\section{Realization of a DF Subspace Using Triangulation}

As a starting point we consider a necessary and sufficient condition for DF
behavior in the projected subspace. This can be determined from the fact
that the interaction part of the intermediate operator $\Theta \left(
t\right) $\ is zero or that the evolution of the projected (reduced) state
is independent of the interaction part of the Liouvillian. That is 
\begin{equation}
L_{0}^{int}+\left( L_{1}^{S}\left( t\right) +L_{1}^{B}+L_{1}^{int}\right)
C\left( t\right) =0.  \label{eqn3a}
\end{equation}%
\ If the constructed projectors $P_{\nu }$ and $Q_{\nu }$ are triangular
with respect to $L_{1}^{S}\left( t\right) +L_{1}^{B}+L_{1}^{int}$ (i.e., $%
P_{\nu }L_{int}P_{\mu }\neq 0$, $P_{\nu }\left( L_{1}^{S}\left( t\right)
+L_{1}^{B}\right) P_{\mu }\neq 0$ for $\nu <\mu $ (up-triangular) or $\nu
>\mu $ (down-triangular)), then $L_{0}^{int}+\left( L_{1}^{S}\left( t\right)
+L_{1}^{B}+L_{1}^{int}\right) C\left( t\right) $ = $L_{0}^{int}+\sum_{\nu
}P_{\nu }\left( L_{1}^{S}\left( t\right) +L_{1}^{B}+L_{1}^{int}\right)
Q_{\nu }\frac{1}{E_{\nu }\left( t\right) -Q_{\nu }L\left( t\right) Q_{\nu }}%
Q_{\nu }\left( L_{1}^{S}\left( t\right) +L_{1}^{B}+L_{1}^{int}\right) P_{\nu
}$ = $0$. This enables the evolution of the reduced projected density
operator to be independent of the interaction part of the total Liouvillian
in the projected Liouville subspace,%
\begin{equation}
\rho ^{proj,S}\left( t\right) =\sum_{\nu }Tr_{B}\left[ \left( P_{\nu
}+D_{\nu }\left( t\right) \right) \rho \left( t\right) \right] =\sum_{\nu
}Tr_{B}\left( e^{-i\int_{t_{0}}^{t}dt^{\prime }l_{\nu }^{\Theta _{0}}\left(
t^{\prime }\right) }P_{\nu }\rho \left( t_{0}\right) \right) ,  \label{eqn3b}
\end{equation}%
where $l_{\nu }^{\Theta _{0}}\left( t\right) $ is a $\nu $th eigenvalue of $%
\Theta _{0}\left( t\right) $ in Liouvillian representation. The
corresponding mixed-state fidelity is equal to $1$ (i.e., $F\left( t\right) $
= $Tr_{S}\left\{ \rho ^{proj,S}\left( 0\right) \sum_{\nu }Tr_{B}\left(
e^{-i\int_{t_{0}}^{t}dt^{\prime }l_{\nu }^{\Theta _{0}}\left( t^{\prime
}\right) }P_{\nu }\rho \left( t_{0}\right) \right) \right\} $ $=$ $1)$. This
shows that there is no decoherence introduced by the environment in the
projected subspace, although the total system is subject to decoherence
introduced by the environment. The formulation for this DF projected
subspace is exact (i.e., there are no approximations including the
Born-Markov approximation). The method is general. Thus for any combined
system, one may construct a triangular basis for the total Hamiltonian to
span a projected subspace in which the evolution of the projected density
operator is independent of the interaction part of the total Hamiltonian,
with a fidelity of $1$. Furthermore, one can construct a partial triangular
basis for the computing system or bath, by choosing a partial diagonal basis
for the left part of total system to form a DF projected subspace, based on
the subdynamical Liouville equation.{\em \ }Thus triangulation procedure
places no restriction on the type of coupling between the system and the
environment (i.e., Markovian or collective decoherence).

\section{An Example}

To illustrate the above method, we revisit the example of the two qubit
quantum computing system $S$, consisting of spins ${\bf S}_{1}$ and ${\bf S}%
_{2}$. For example, this includes the two electrons around two $^{31}P$
confined in a germanium/silicon heterostructures in an electron
spin-resonance transistor$^{\left[ 1\right] }$, or two electrons confined in
two quantum dots$^{\left[ 2\right] }$. Ignoring the influence of
environment, the Hamiltonian can be written using the Heisenberg model as $%
H_{S}\left( t\right) =J\left( t\right) {\bf S}_{1}\cdot {\bf S}_{2}$, where $%
J\left( t\right) $ is the time-dependent exchange coupling parameter
determined by the specific model considerations. The coupling to the
environment is assumed to be described by a Caldeira-Leggett-type model,
consisting of a set of harmonic oscillators coupled linearly to $S$ by $%
H_{int}=\lambda \sum_{k=1}^{2}{\bf \sigma }_{k}\cdot {\bf b}_{k}$, where $%
b_{k}^{j}=\sum_{\alpha }g_{\alpha }^{kj}\left( a_{\alpha ,kj}+a_{\alpha
,kj}^{+}\right) $ is a fluctuation quantum field. The Hamiltonian of bath is
given by $H_{B}=\sum_{\alpha }\omega _{\alpha }^{kj}a_{\alpha
,kj}^{+}a_{\alpha ,kj}$, where $a_{\alpha ,kj}^{+}$ ($a_{\alpha ,kj}$) are
bosonic creation (annihilation) operators with $j=x,y,z$ and $\omega
_{\alpha }^{kj}$ are the corresponding frequencies with spectral
distribution function $J_{kj}\left( \omega \right) =\pi \sum_{\alpha }\left(
g_{\alpha }^{kj}\right) ^{2}\delta \left( \omega -\omega _{\alpha }\right) $ 
$^{\left[ 28\right] }$.

We show how to triangulate the partial basis of the bath since, in
principle, triangulating the partial basis of the system follows the same
approach. The key point is that, in general, the triangular property of $L$
results in $P_{\nu }LQ_{\nu }C_{\nu }P_{\nu }$ being in the SKE.

The matrix for $H_{B}+H_{int}$, with respect to a $\alpha $th element of the
basis of $B$, in a repeat subspace, $\left( \left| n_{\alpha
,k}^{j}\right\rangle \text{, }\left| n_{\alpha ,k}^{j}+1\right\rangle \text{%
, }\left\langle n_{\alpha ,k}^{j}\right| \text{, }\left\langle n_{\alpha
,k}^{j}+1\right| \right) $, can be written as 
\begin{equation}
M\equiv \left( 
\begin{array}{cc}
\omega _{\alpha }^{kj}n_{\alpha ,k}^{j} & g_{\alpha }^{kj}\sqrt{n_{\alpha
,k}^{j}+1} \\ 
g_{\alpha }^{kj}\sqrt{n_{\alpha ,k}^{j}+1} & \omega _{\alpha }^{kj}\left(
n_{\alpha ,k}^{j}+1\right)%
\end{array}%
\right) .  \label{eqnb1}
\end{equation}%
Using a similarity transformation to up-triangulate $M$, 
\begin{eqnarray}
M^{tri} &=&\frac{1}{ad-bc}\left( 
\begin{array}{cc}
d & -b \\ 
-c & a%
\end{array}%
\right) \left( 
\begin{array}{cc}
\omega _{\alpha }^{kj}n_{\alpha ,k}^{j} & g_{\alpha }^{kj}\sqrt{n_{\alpha
,k}^{j}+1} \\ 
g_{\alpha }^{kj}\sqrt{n_{\alpha ,k}^{j}+1} & \omega _{\alpha }^{kj}\left(
n_{\alpha ,k}^{j}+1\right)%
\end{array}%
\right) \left( 
\begin{array}{cc}
a & b \\ 
c & d%
\end{array}%
\right)  \label{eqn4aa} \\
&=&\ \ \frac{1}{ad-bc}\left( 
\begin{array}{cc}
ad\omega _{\alpha }^{kj}n_{\alpha ,k}^{j} & \left( d^{2}-b^{2}\right)
g_{\alpha }^{kj}\sqrt{n_{\alpha ,k}^{j}+1}-bd\omega _{\alpha }^{kj} \\ 
0 & -bc\omega _{\alpha }^{kj}\left( n_{\alpha ,k}^{j}+2\right) +ad\omega
_{\alpha }^{kj}\left( n_{\alpha ,k}^{j}+1\right)%
\end{array}%
\right) ,  \nonumber
\end{eqnarray}%
we have 
\begin{eqnarray}
-bc\omega _{\alpha }^{kj}\left( n_{\alpha ,k}^{j}+1\right) +\left(
dc-ab\right) g_{\alpha }^{kj}\sqrt{n_{\alpha ,k}^{j}+1} &=&0,  \nonumber \\
\left( -c^{2}+a^{2}\right) g_{\alpha }^{kj}\sqrt{n_{\alpha ,k}^{j}+1}%
+ac\omega _{\alpha }^{kj} &=&0,  \nonumber
\end{eqnarray}%
which gives 
\begin{eqnarray}
a &=&\left( \frac{-\omega _{\alpha }^{kj}}{2g_{\alpha }^{kj}\sqrt{n_{\alpha
,k}^{j}+1}}\pm \sqrt{\left( \frac{\omega _{\alpha }^{kj}}{2g_{\alpha }^{kj}}%
\right) ^{2}\frac{1}{n_{\alpha ,k}^{j}+1}+1}\right) c\equiv \gamma _{\alpha
,kj}c,  \label{eqn4bb} \\
d &=&\frac{\omega _{\alpha }^{kj}\sqrt{n_{\alpha ,k}^{j}+1}+\gamma _{\alpha
,kj}g_{\alpha }^{kj}}{g_{\alpha }^{kj}}b\equiv \zeta _{\alpha ,kj}b. 
\nonumber
\end{eqnarray}%
This determines the triangular basis for $H_{B}+H_{int}$ in the repeat
subspace as $\left( \left| \phi _{n}\otimes \left\{ n_{\alpha ,k}^{j,\pm
}\right\} \right\rangle \text{, }\left\langle \left\{ n_{\alpha ,k}^{j,\pm
}\right\} \otimes \phi _{n}\right| \right) $. For example, by choosing $c=1$
and $b=-1$ we have 
\begin{equation}
M^{tri}=\left( 
\begin{array}{cc}
-\gamma _{\alpha ,kj}\zeta _{\alpha ,kj}\omega _{\alpha }^{kj}n_{\alpha
,k}^{j} & \left( \zeta _{\alpha ,kj}^{2}-1\right) g_{\alpha }^{kj}\sqrt{%
n_{\alpha ,k}^{j}+1}-\zeta _{\alpha ,kj}\omega _{\alpha }^{kj} \\ 
0 & \omega _{\alpha }^{kj}\left( n_{\alpha ,k}^{j}+2\right) -\gamma _{\alpha
,kj}\zeta _{\alpha ,kj}\omega _{\alpha }^{kj}\left( n_{\alpha
,k}^{j}+1\right)%
\end{array}%
\right) ,  \label{eqnb3}
\end{equation}%
defining, 
\begin{equation}
\left( 
\begin{array}{c}
\left| n_{\alpha ,k}^{j,+}\right\rangle \\ 
\left| n_{\alpha ,k}^{j,-}\right\rangle%
\end{array}%
\right) =\left( 
\begin{array}{c}
\gamma _{\alpha ,kj}\left| n_{\alpha ,k}^{j}\right\rangle -\left| n_{\alpha
,k}^{j}+1\right\rangle \\ 
\left| n_{\alpha ,k}^{j}\right\rangle -\zeta _{\alpha ,kj}\left| n_{\alpha
,k}^{j}+1\right\rangle%
\end{array}%
\right) ,  \label{eqnb4}
\end{equation}%
and%
\begin{equation}
\left( 
\begin{array}{c}
\left\langle n_{\alpha ,k}^{j,+}\right| \\ 
\left\langle n_{\alpha ,k}^{j,-}\right|%
\end{array}%
\right) =\frac{1}{-\gamma _{\alpha ,kj}\zeta _{\alpha ,kj}+1}\left( 
\begin{array}{c}
-\zeta _{\alpha ,kj}\left\langle n_{\alpha ,k}^{j}\right| -\text{ }%
\left\langle n_{\alpha ,k}^{j}+1\right| \\ 
\left\langle n_{\alpha ,k}^{j}\right| +\gamma _{\alpha ,kj}\left\langle
n_{\alpha ,k}^{j}+1\right|%
\end{array}%
\right)  \label{eqnb5}
\end{equation}%
with $\left\{ n_{\alpha ,k}^{j,\pm }\right\} $ = $\left( n_{1,k}^{j,\pm
}\cdots n_{\alpha ,k}^{j,\pm }\cdots \right) $. Then we can construct a
triangular basis for the total Liouvillian $L\left( t\right) $ by \{$\left|
\left| \phi _{n}\otimes \left\{ n_{\alpha ,k}^{j,\pm }\right\} \right\rangle
\left\langle \left\{ n_{\beta ,k}^{j,\pm }\right\} \otimes \phi _{m}\right|
\right) $, $\left( \left| \phi _{m}\otimes \left\{ n_{\beta ,k}^{j,\pm
}\right\} \right\rangle \left\langle \left\{ n_{\alpha ,k}^{j,\pm }\right\}
\otimes \phi _{n}\right| \right| $\}. This gives triangulation of matrix of $%
L_{B}+L_{int}$. By means of the new triangular basis one can define the
projectors as $P_{\nu }\equiv \left| \left| \phi _{n}\otimes \left\{
n_{\alpha ,k}^{j,\pm }\right\} \right\rangle \left\langle \left\{ n_{\beta
,k}^{j,\mp }\right\} \otimes \phi _{m}\right| \right) \left( \left| \phi
_{m}\otimes \left\{ n_{\beta ,k}^{j,\mp }\right\} \right\rangle \left\langle
\left\{ n_{\alpha ,k}^{j,\pm }\right\} \otimes \phi _{n}\right| \right| $
and $Q_{\nu }\equiv 1-P_{\nu }$. Taking into account the definitions of the
creation and destruction operators, one has $P_{\nu }\left(
L_{1}^{B}+L_{1}^{int}\right) P_{\mu }C_{\nu }\left( t\right) P_{\nu }\sim
P_{\nu }\left( L_{1}^{B}+L_{1}^{int}\right) P_{\mu }\left(
L_{1}^{B}+L_{1}^{int}\right) P_{\nu }\longrightarrow 0$ for $\mu >\nu $
(up-triangular property). Therefore one finds the DF condition: $%
L_{0}^{int}+\left( L_{1}^{B}+L_{1}^{int}\right) C\left( t\right) $ = $0$
with $\Theta \left( t\right) $ = $\Theta _{0}\left( t\right) $ = $%
L_{0}^{S}+L_{0}^{B}$. Under this DF condition, the SKE reduces to $i\frac{%
\partial \left| \rho ^{proj}\left( t\right) \right) }{\partial t}$ = $\Theta
_{0}\left( t\right) \rho ^{proj}\left( t\right) $. The evolution of the
reduced projected state, in the initial decoupling condition, becomes to Eq.(%
\ref{eqn3b}). This shows that $\rho _{S}^{proj}\left( t\right) $ is
independent on $L_{int}$. Therefore there is no decoherence introduced by $%
L_{int}$ in the projected subsystem although the total system is subject to
the decoherence introduced by $L_{int}$. In this projected subsystem, the
quantum Control-Not logic operation is still given by a sequence of
operations and the swap operator $U_{sw}$ remains invariant before and after
the interaction, and is given by 
\begin{eqnarray}
Tr_{B}\widehat{T}e^{-i\int_{0}^{\tau _{s}}\Theta \left( \tau \right) d\tau
}\rho ^{proj}\left( 0\right) &=&Tr_{B}\widehat{T}e^{-i\int_{0}^{\tau
_{s}}L_{0}^{S}\left( \tau \right) d\tau }\rho ^{proj}\left( 0\right)
\label{eqnn} \\
&=&Tr_{B}\left( \widehat{T}e^{-i\int_{0}^{\tau _{s}}H_{0}^{S}\left( \tau
\right) d\tau }\rho ^{proj}\left( 0\right) \widehat{T}e^{i\int_{0}^{\tau
_{s}}H_{0}^{S}\left( \tau \right) d\tau }\right)  \nonumber \\
&=&Tr_{B}\left( U_{sw}\rho ^{proj}\left( 0\right) U_{sw}^{-1}\right) , 
\nonumber
\end{eqnarray}%
where%
\begin{equation}
U_{sw}=\sum_{\alpha }\left( \sum_{n=1}^{3}e^{-i\frac{\pi }{4}}\left| \phi
_{n}\otimes {\bf 1}_{\alpha }\right\rangle \left\langle {\bf 1}_{\alpha
}\otimes \phi _{n}\right| +e^{i\frac{3\pi }{4}}\left| \phi _{4}\otimes {\bf 1%
}_{\alpha }\right\rangle \left\langle {\bf 1}_{\alpha }\otimes \phi
_{4}\right| \right) .  \label{eqss}
\end{equation}

The following table clarifies the difference between the total space and the
projected subspace constructed by the triangulation:

\[
\begin{tabular}{|l|l|l|l|}
\hline
& $Initial\text{ }State\text{ }$ & $Evolutionary\text{ }State$ & $Fidelity$
\\ \hline
$Subspace$ & $Tr_{B}P_{\nu }\rho _{S}\left( 0\right) \otimes \rho _{B}\left(
0\right) $ & $Tr_{B}\widehat{T}e^{-i\int L_{0}^{S}\left( t\right) dt}P_{\nu
}\rho _{S}\left( 0\right) \otimes \rho _{B}\left( 0\right) $ & $f=1$ \\ 
\hline
$Total\text{ }Space$ & $Tr_{B}\rho _{S}\left( 0\right) \otimes \rho
_{B}\left( 0\right) $ & $Tr_{B}\widehat{T}e^{-i\int \left( L_{0}^{S}\left(
t\right) +L_{1}^{B}+L_{1}^{int}\left( t\right) \right) dt}\rho _{S}\left(
0\right) \otimes \rho _{B}\left( 0\right) $ & $0<f<1$ \\ \hline
\end{tabular}%
\]%
If one assumes that in the Schr\"{o}dinger picture, an initial state of $S$
is $\phi _{S}\left( 0\right) $ = $\frac{1}{2}\left( \left| \uparrow
\downarrow \right\rangle +\left| \downarrow \uparrow \right\rangle \right) -%
\frac{1}{2}\left( \left| \uparrow \uparrow \right\rangle +\left| \downarrow
\downarrow \right\rangle \right) $, then by choosing a projector $P_{\nu }$
as $\left| \frac{1}{2}\left( \left| \uparrow \downarrow \right\rangle
+\left| \downarrow \uparrow \right\rangle \right) \otimes \left\{ n_{\alpha
,k}^{j,\pm }\right\} \right) \left( \left\{ n_{\alpha ,k}^{j,\pm }\right\}
\otimes \frac{1}{2}\left( \left| \uparrow \downarrow \right\rangle +\left|
\downarrow \uparrow \right\rangle \right) \right| $, after taking into
account the obtained triangular basis, an initial state in the DF $P_{\nu }$%
-projected subspace is given by $Tr_{B}P_{\nu }\phi _{S}\left( 0\right) $ = $%
\frac{1}{2}\left( \left| \uparrow \downarrow \right\rangle +\left|
\downarrow \uparrow \right\rangle \right) $ under the initial decoupling
condition. This correspondence is realized by the projection: $\phi
_{S}\left( 0\right) \longrightarrow P_{\nu }^{S}\phi _{S}\left( 0\right) $.
The advantage of using $P_{\nu }^{S}\phi _{S}\left( 0\right) $ compared with 
$\phi _{S}\left( 0\right) $, is that the evolution of $P_{\nu }^{S}\phi
_{S}\left( 0\right) $ is independent of the interaction between $S$ and $B$
in the projected subspace. Therefore, one can encode information onto $%
P_{\nu }^{S}\phi _{S}\left( 0\right) $ to perform quantum computing (or
quantum communication), protecting against decoherence introduced by the
interaction between $S$ and $B$.

Finally, it should be noted that it is not necessary that the evolution
operator $U_{\Theta }$ in the projected subspace is unitary, because a
quantum computing system projected in a DF subspace may be an open quantum
system, obeying the semigroup evolution rules$^{\left[ 17\right] }$. In this
open system, self-adjoint operators and unitary evolution groups are not
intrinsically necessary to govern quantum computation. Quantum computation
can then be performed in a more general functional space, such as RHS,
rather than just Hilbert space. \ The projected state $\psi ^{proj}$ may
exist in the test space $\Phi ^{proj}$, which is a dense subspace of the
Hilbert space ${\cal H}^{proj}$ constructed by $\sum_{\nu }P_{\nu }\Pi _{\nu
}{\cal H}$, representing the physical states which can be prepared in an
actual experiment. Its adjoint $\widetilde{\psi }^{proj}$ lies in the dual
space $\left( \Phi ^{proj}\right) ^{\times }$, representing a procedure that
associates with each state a number, while preserving the linear structure
which results from the superposition principle, i.e., the triplet structure $%
\Phi ^{proj}\subset {\cal H}^{proj}\subset $ $\left( \Phi ^{proj}\right)
^{\times }$ $^{\left[ 23\right] }$. This is a RHS structure which
facilitates describing irreversible processes like decoherence and
dissipation due to interaction with the environment. In this space the
evolution of the states are permitted to be time asymmetric, providing a
framework for describing the irreversibility of practical open systems. This
irreversibility does not change quantum reversible logical operations to
quantum irreversible logical operation in the quantum universal
Controlled-Not logical gate. To appreciate this one must distinguish between
irreversibility of a quantum logical operation, introduced by the structure
of logical gate, and irreversibility of the process induced by interactions
with the environment. Reversible computation means reversible logical
operations on the structure of the logical gate. In this sense, quantum
computing in RHS is compatible with reversible quantum logical operations
and permits computing any reversible function, although irreversible
processes do in fact exist.

\section{Conclusion}

In conclusion, a subdynamics based formulation in the Schr\"{o}dinger
picture was presented for an open quantum system. Based on the subdynamical
kinetic equation for an open quantum system, a proposal for quantum
computing in the projected subspaces is developed. The eigenvectors of the
intermediate operator in this subspace were shown to remain invariant before
and after interaction with the environment, while the eigenvalues of the
intermediate operator in this subspace change. The fidelity of mixed states
in the projected subspace is $1$ which means that the constructed projected
subspace is definitely DF. This reveals a universal property for any system
plus reservoir: one can construct a DF projected subspace by using
eigenprojectors of the free part of the Hamiltonian in which the encoded
states are projected states, which are themselves determined by relevant
formulae of subdynamics. On the other hand, changes of the eigenvalues after
interaction may introduce a type of unitary error in the ideal swap
operator. This sort of error can be cancelled by adjusting the interaction
coupling time between two spins in the subspace, since the eigenvectors
remain invariant, although decoherence exists in the total space for the
total system. Finally, the general case for completely DF behavior in the
projected subspaces was discussed, and it was shown that using a general DF
condition (i.e., the second term of the subdynamics kinetic equation is
zero), one can find a condition to allow the constructed projected subspace
to be DF. This reveals that this condition is a necessary and sufficient
condition for constructing a DF projected subspace. We wish to emphasize two
points here: (1) \bigskip the constructed projected subspace is spanned by a
set of $\left\{ \rho _{\nu }^{proj}\left( t\right) \right\} $, which is
closed with respect to the intermediate operator. Indeed, $\Theta \left(
t\right) \rho ^{proj}\left( t\right) $ = $\sum_{\nu }P_{\nu }L\left(
t\right) \left( P_{\nu }+C_{\nu }\left( t\right) \right) \Pi _{\nu }\left(
t\right) \rho \left( t\right) $ = $\sum_{\nu }l_{\nu }^{\Theta }\left(
t\right) \rho _{\nu }^{proj}\left( t\right) $, where $l_{\nu }^{\Theta
}\left( t\right) $ is a $\nu $th eigenvalue of $\Theta \left( t\right) $ in
Liouvillian representation. (2) The Born-Markov assumption and various other
types of restrictions for DF behavior does not need to be made, indeed as we
show a general approach can be used.

{\LARGE \ }

{\LARGE \ }

{\LARGE \ }

\bigskip {\bf ACKNOWLEDGEMENTS}

{\LARGE \ }

We gratefully acknowledge financial support from grants from NSERC, MITACS,
CIPI, MMO, CITO and China State Key Projects of Basic Research and Natural
Science foundation (G1999064509, N$_{0}$ 79970121, 60072032).

\end{document}